\documentstyle[11pt,newpasp,twoside,epsf]{article}
\def\edcomment#1{\iffalse\marginpar{\raggedright\sl#1\/}\else\relax\fi}
\reversemarginpar

\newcommand{\fig}[6]{
    \protect\centerline{
    \epsfxsize=#1\epsffile[#2 #3 #4 #5]{#6}
    }}

\begin{document}
\title{Testing the Unified Model with an Infrared Selected Sample of Seyferts}
\author{H. R. Schmitt and J. S. Ulvestad}
\affil{National Radio Astronomy Observatory, P.O. Box 0, Socorro, NM87801}
\author{R. R. J. Antonucci}
\affil{University of California, Santa Barbara, Santa Barbara, CA93106}
\author{C. J. Clarke and J. E. Pringle}
\affil{Institute of Astronomy, Madingley Road, Cambridge CB3 0HA, England}
\author{A. L. Kinney}
\affil{NASA Headquarters, 300 E St., Washington, DC20546}

\begin{abstract}

We present a series of statistical tests using homogeneous data and
measurements for a sample of Seyfert galaxies. These galaxies were
selected from mostly isotropic properties, their far infrared fluxes
and warm infrared colors, which provide a considerable
advantage over the criteria used by most investigators in the past,
like ultraviolet excess. Our results provide strong support for a
Unified Model in which Seyferts 2's contain a torus seen more edge-on
than in Seyferts 1's and show that previous results showing
the opposite were most likely due to selection effects.

\end{abstract}

\section{Introduction}

The Unified Scheme is based on the idea that the nucleus
is surrounded by a dusty molecular torus, with orientation angle
being the parameter which determines whether an AGN is perceived by
observers as a Seyfert~1 or as a Seyfert~2 (Antonucci 1993). This scenario
is supported by the detection of polarized broad emission lines in Seyfert 2's
(Antonucci \& Miller 1985) and the collimated escape of radiation from
nuclear region, detected as Narrow Line Regions with conical shapes
in Seyfert 2 galaxies (Pogge 1989) and jet like radio emission
(Ulvestad \& Wilson 1989).

It is now accepted that the Unified Model applies to a large fraction of
Seyfert galaxies and is correct to first order. However, some
observational results claim intrinsic statistical differences between
Seyfert 1's and Seyfert 2's. Malkan et al. (1998) found that
Seyfert 1's usually reside in earlier type host galaxies compared to
Seyfert 2's, while Laurikainen \& Salo (1995) and Dultzin-Hacian et al. (1999)
found a higher percentage of companions around  Seyfert 2's than in
Seyfert 1's.

We believe that these differences between Seyfert 1's and Seyfert 2's
are mostly due to the way these papers selected their samples.
In order to be able to make a fair comparison
between Seyfert 1's and Seyfert 2's and correctly address problems
related to the Unified Model, it is necessary to use a sample selected
from an isotropic property, believed to be independent of the
orientation of the torus relative to the line of sight.

\section{Sample and Data}

One of the best ways to select an isotropic sample is based on the far
infrared properties of the galaxies.  According to the Pier \& Krolik
(1992) torus models, the circumnuclear torus radiates nearly
isotropically at 60$\mu$m, so a sample selected in this way should be
relatively free from selection effects.

Our sample was extracted from the survey of warm Seyfert galaxies defined
by de Grijp et al. (1992), which was selected based on the strength of
their IRAS 25$\mu$m and 60$\mu$m fluxes and warm infrared colors.
We selected all Seyfert galaxies with z$\leq$0.031, which gives a total
of 29 Seyfert 1's and 59 Seyfert 2's.
We have to point out, however, that this sample is not complete, since any
Seyfert galaxy with infrared colors cooler than our criteria or for which
there is no good quality IRAS data, are missed.

Another important point in this study is the use of high quality radio and
optical images, obtained and measured homogeneously. We use broadband
B and I images obtained for all the galaxies by Schmitt \& Kinney (2000),
which will be used to determine host galaxy inclinations and, in some
cases, morphological types. We also used VLA A-configuration 3.6cm continuum
images for 75 of the galaxies (Schmitt et al. 2001a),
which will be used to determine the 3.6cm radio powers and the extent
of the radio emission in these galaxies.

\section{Infrared and Radio Luminosities}

The left panel of Figure 1 presents the distribution of 60$\mu$m luminosities
for Seyfert 1's and Seyfert 2's. We can see that they
have very similar distributions, with the Kolmogorov-Smirnov test (KS-test)
giving a 45\% chance that two samples drawn from the same parent population
would differ this much. This result is expected since the galaxies were
selected by their infrared properties and shows that this measurement is
isotropic.

The distribution of 3.6cm radio powers is presented on the right panel of
Figure 1. Both Seyfert types have similar radio power
distributions, as expected from the Unified Model, with the KS test showing
that two samples selected from the same parent population would differ this
much 11\% of the time.

\begin{figure}
    \fig{11cm}{80}{220}{550}{460}{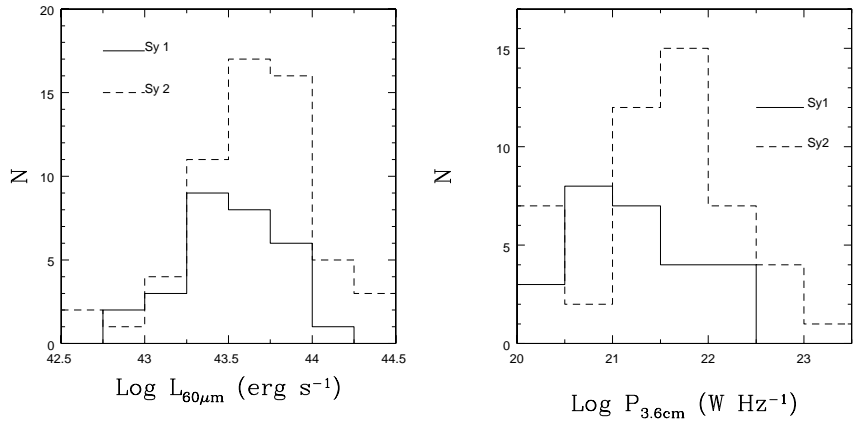}
\caption{Left: histogram of the 60$\mu$m luminosity; right: histogram of the
radio continuum 3.6cm power. Seyfert 1's and Seyfert 2's are represented by
solid and dashed lines, respectively.}
\end{figure}

\section{Host Galaxy Properties}

The left panel of Figure 2 shows the observed distribution of the ratio between 
the host galaxies minor to major axis lengths (b/a). The observed distribution
has a deficit of Seyfert 1 galaxies with b/a$<$0.5, while in Seyfert 2's
this does not happen. The lack of galaxies with b/a$<$0.2 is
due to the thickness of the disk. Comparing the b/a distribution for
Seyfert 1's and Seyfert 2's we find that they are significantly different,
with the KS test giving the probability that
two samples drawn from the same parent population would differ this much
only 4.7\% of the time.

This result is similar to the one found by Keel (1980), who was
the first to discover a deficiency of edge-on Seyfert 1
galaxies (see also Lawrence \& Elvis 1982 and Maiolino \& Rieke 1995).
Although this result is in principle not expected from the Unified Model,
it does not necessarily contradict it. The papers cited above suggested that,
in the case of edge-on Seyfert galaxies, the gas and dust along the host
galaxy disk can block the direct view of the Broad Line Region, thus
leading to a classification as a Seyfert 2 galaxy.
 
The comparison between the Morphological Types of the host galaxies of
Seyfert 1's and Seyfert 2's is shown on the right panel of Figure 2.
The two distributions are very similar, with the KS test showing
that two samples selected from the same parent population would differ this
much 80\% of the time. 

\begin{figure}
    \fig{11cm}{80}{220}{550}{460}{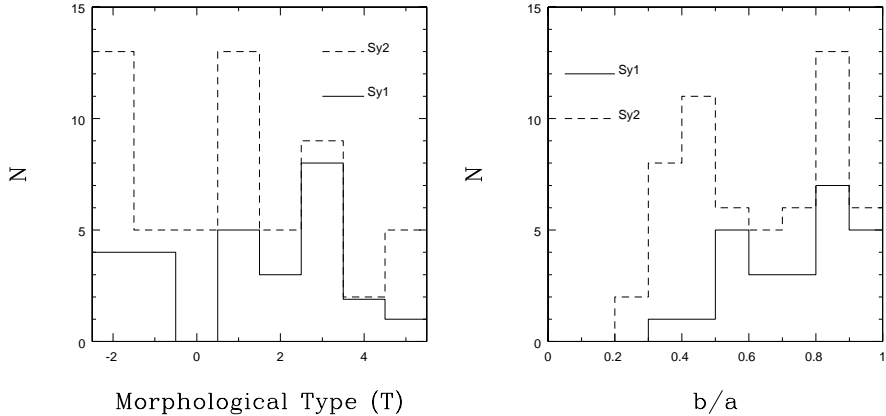}
\caption{Histograms of morphological types, left, and the ratio between
the semiminor and semimajor axes of the host galaxies, right.}
\end{figure}

\section{Radio Sizes}

The distributions of the logarithm of the extension of the 3.6cm radio
emission in Seyfert 1's and Seyfert 2's is presented in Figure 3.
Only 9 out of 26 Seyfert 1's (35\%) show extended emission, and the rest are
unresolved, as indicated by arrows in the figure. In the case of Seyfert 2's,
28 out of 48 galaxies (58\%) have extended radio emission.

Given the fact that 50\% of the galaxies in our sample are unresolved,
we had to use survival analysis to compare the two distributions.
The mean and standard deviations of the extension of the radio emission
were calculated using the Kaplan-Meier estimator, which gave
148$\pm$65 pc and 348$\pm$97 pc for Seyfert 1's and Seyfert 2's, respectively.
We also compared whether Seyfert 1's and Seyfert 2's have similar
distributions of the extension of the radio emission, using the
Gehan-Wilcoxon test. We obtained a probability
of 4.2\% that the two samples are drawn from the same parent population.
This confirms that Seyfert 1's have smaller extended emission, as
predicted by the Model.

\begin{figure}
    \fig{7.5cm}{60}{180}{460}{580}{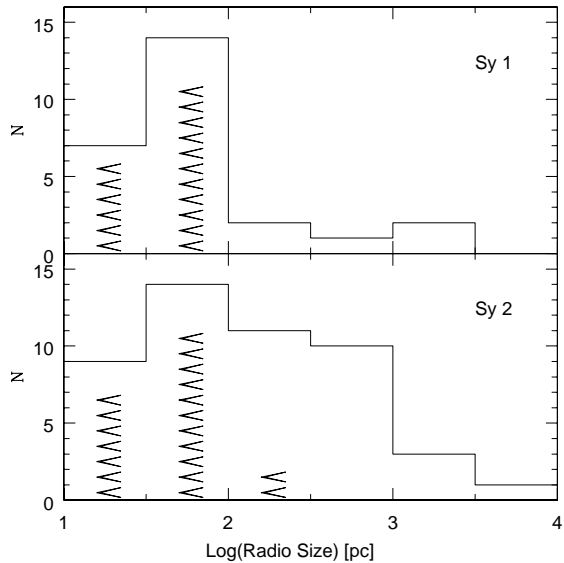}
\caption{The distribution of the logarithm of the extension of the radio
emission in Seyfert 1's and Seyfert 2's. The histograms represent
the total number of galaxies in each bin, adding those with detected extended
emission and upper limits, represent by arrows.}
\end{figure}

\section{Frequency of Companions}

Several mechanisms have been suggested to explain how
to transport gas from the disk of a spiral
galaxy to its nucleus, like interactions (Gunn 1979; Hernquist 1989)
or bars (Schwartz 1981).
The influence of interactions on the fueling of AGN has been the topic
of several papers, but so far there is no consensus about this subject.
Dahari (1984) and Rafanelli et al. (1995), among others, found Seyferts to have
an excess of companions relative to normal galaxies. On the other hand,
Fuentes-Williams \& Stocke (1988) and Bushouse (1986) found that there is
no detectable difference in the environments of Seyfert and normal galaxies.
An intriguing result was obtained by Laurikainen \& Salo (1995) and
Dultzin-Hacyan et al. (1999), who showed that Seyfert 2's have a larger number
of companions when compared to normal galaxies, while Seyfert 1's do not.

We used our broad band images, NED and the Digitized Sky Survey plates (DSS)
to search for companions around our galaxies, and adopted the parameters
used by Rafanelli et al. (1995) to determine if a galaxy has a companions.
A galaxy is considered a companion if its distance
to the galaxy of interest is smaller than 3 times the diameter of that
galaxy (3D), the difference in brightness between them is smaller than 3
magnitudes ($|\Delta m|\leq$ 3 mag) and the radial velocities difference is
smaller than $|c\Delta z|\leq$1000 km s$^{-1}$.

According to these criteria, a total of 25 out of the 88 Seyfert
galaxies in our sample have galaxies with $|\Delta I|$ or
$|\Delta B|\leq3$~mag and closer than 3D from them, which puts
an upper limit of $<28\pm$6\% of possible interacting galaxies in this
sample (the uncertainty is given by Poisson statistics).
Of these 25 galaxies, 9 are Seyfert 1's and 16 are Seyfert 2's,
which gives an upper limit of possible companions of $<31\pm$10\% and
$<27\pm$7\%, respectively. When we consider only the galaxies
which satisfy the brightness, distance and velocity criteria,
we find 17 Seyferts with confirmed companions. Since there is no
information on the radial velocities for 8 of the possible companion
galaxies, we assume that 17 is a lower limit, which gives
that the percentage of Seyferts with companions is
$>19\pm5$\%. Of these 17 galaxies, 7 are Seyfert 1's and 10 are Seyfert 2's,
corresponding to a lower limit of $>24\pm9$\% and $>17\pm5$\% galaxies with
companions, respectively.

The percentage of confirmed companions in our sample is similar to the one
found by Rafanelli et al. (1995) for the CfA sample, and also to the ones
obtained by Schmitt (2001) for the Palomar sample, which is the same as
for galaxies with other activity types. We should also notice
that there is no apparent difference in the upper and lower percentage
of companion galaxies in Seyfert 1's and Seyfert 2's, contradicting the
results obtained by Laurikainen \& Salo (1995) and Dultzin-Hacyan et al.
(1999). An explanation why these paper got to their results is given by
Schmitt et al. (2001b).

\section{Summary}

We presented a series of test to the Unified Model of Seyfert galaxies
based on a sample of galaxies selected by their infrared properties, which
presents several advantages relative to other samples.
The far infrared is a  mostly isotropic property, which is essential
for testing Unified Models, since the sample is unbiased relative to the
orientation of the torus. We also used homogeneous data and measurements.
The detailed results are presented by Schmitt et al. (2001b).

We found that both Seyfert 1's and Seyfert 2's have similar 60$\mu$m
luminosities and 3.6cm radio powers, as well as similar
distributions of morphological types. The comparison between the host
galaxy inclinations shows that there is a deficiency of Seyfert 1's
in edge-on galaxies, which was known from previous studies and
apparently contradicts the Unified Model. However, the model can be reconciled
with the observations if we assume that Seyfert 1's observed edge-on will
have their nucleus hidden by gas and dust in the galaxy disk.
The extension of the radio emission in Seyfert 1's is, on average,
smaller than in Seyfert 2's, as expected from the model.
We also show that there is a similar percentage of Seyfert 1's and Seyfert 2's
with companions.

These results, taken together, give strong support to the Unified Model. This
indicates that previous results, which found differences in isotropic
properties of Seyfert 1's and Seyfert 2's, were most likely due to selection
effects.

\acknowledgements

Support for this work was provided by NASA grant AR-8383.01-97A.
The National Radio Astronomy Observatory is a facility
of the National Science Foundation operated under cooperative agreement
by Associated Universities, Inc.


\begin{references}
Antonucci, R. R. J. 1993, ARA\&A, 31, 473\\
Antonucci, R. R. J. \& Miller, J. S. 1985, ApJ, 297, 621\\
Bushouse, H. A. 1986, AJ, 91, 255\\
Dahari, O. 1984, AJ, 89, 966\\ 
de Grijp, M. H. K. et al. 1992, A\&AS, 96, 389\\
Dultzin-Hacyan, D. et al. 1999, ApJ, 513, L111\\
Fuentes-Williams, T. \& Stocke, J. T. 1988, AJ, 96, 1235\\
Gunn, J. 1979, in Active Galactic Nuclei, edited by C. Hazard \& S. Mitton,
(Cambridge University Press, Cambridge), p.213\\
Hernquist, L. 1989, Nature, 640, 687\\
Keel, W. C. 1980, AJ, 85, 198\\
Kinney, A. L. et al. 2000, ApJ, 537, 152\\
Laurikainen, E. \& Salo, H. 1995, A\&A, 293, 683\\
Lawrence, A. \& Elvis, M. 1982, ApJ, 256, 410\\
Maiolino, R. \& Rieke, G. H. 1995, ApJ, 454, 95\\
Malkan, M. A., Gorjian, V. \& Tam, R. 1998, ApJS, 117, 25\\
Pier, E. A. \& Krolik, J. H. 1992, ApJ, 401, 99\\
Pogge, R. W. 1989, ApJ, 345, 730\\
Rafanelli, P., Violato, M. \& Baruffolo, A. 1995, AJ, 109, 1546\\
Schmitt, H. R. 2001, AJ, in press\\
Schmitt, H. R. \& Kinney, A. L. 1996, ApJ, 463, 498\\
Schmitt, H. R. et al. 2001a, ApJS, 132, 199\\
Schmitt, H. R. et al. 2001b, ApJ, 555, 663\\ 
Schwartz, M. 1981, ApJ, 247, 77\\
Ulvestad, J. S., \& Wilson, A. S. 1989, ApJ, 343, 659\\
\end{references}
\end{document}